\documentstyle[epsfig,aps]{revtex}
\begin{document}
\draft
\title{A Raman Investigation of $PbZr_{0.94}Ti_{0.06}O_3$ Ceramics Under
High-Pressures}
\author{A.G. Souza Filho$^{a,\dag}$, P.T.C. Freire$^{a,*}$, J.M. Sasaki$^a$,
I. Guedes$^a$, J. Mendes Filho$^a$, F.E.A. Melo$^a$, E.B.
Ara\'ujo$^b$, and J.A. Eiras$^b$}
\address{$^a$Departamento de F\'{\i}sica, Universidade Federal do Cear\'a,
Caixa Postal 6030\\ Campus do Pici, 60455-970 Fortaleza, Cear\'a,
Brazil, $^\dag$ e-mail:agsf@fisica.ufc.br}
\address{$^b$Universidade Federal de S\~ao Carlos, Departamento de F\'{\i}sica,
Grupo de Cer\^amicas Ferroel\'etricas, Caixa Postal 676, 13565-670
S\~ao Carlos SP - Brazil} \maketitle

\begin{abstract}

We have investigated the behavior of the $PbZr_{0.94}Ti_{0.06}O_3$
Raman spectra as a function of hydrostatic pressures. The new
structural phases were identified based on previous works
performed in PZT system with different Ti concentrations at room-
and high pressures. We showed that $PbZr_{0.94}Ti_{0.06}O_3$
exhibits a rich phase sequence up to 3.7 $GPa$:
$rhombohedral(LT)\stackrel{0.3 GPa}{\rightarrow} orthorhombic(I)
\stackrel{2.9 GPa}{\rightarrow}orthorhombic(I')$. This sequence is
different from that exhibited by $PbZr_{0.90}Ti_{0.10}O_3$
suggesting a very interesting concentration-pressure phase diagram
for rich Zr PZT system.
\end{abstract}
\bigskip
\vspace{0.2cm} \centerline{Keywords A. insulators, D. phase
transitions, E. inelastic light scattering, E. high pressure}
\bigskip
\section{Introduction}
\bigskip
Lead zirconate titanate (PZT) is a solid solution of lead
zirconate and lead titanate with molecular formula
$Pb(Zr_{1-x}Ti_x)O_3$, known by their excellent pyroelectric,
piezoelectric ferroelectric and properties. At room temperature, a
variety of PZT forms (thin films, ceramics and single crystals)
with different $Ti$ concentrations have been widely studied by
many experimental methods such as X-ray scattering \cite{hz,pig} ,
neutron diffraction \cite{kata} , dielectric and pyroelectric
measurements\cite{z,ujma} and Raman
spectroscopy\cite{becker,db,dca,kr,burns,scott,meng}.

Previous works have shown that PZT has several structures at room
temperature depending on the $x$ value. Among these structures PZT
appears as tetragonal $(C_{4v}^1)$,  orthorhombic $(C_{2v}^8)$ and
rhombohedral $(C_{3v}^1$ and $C_{3v}^6)$ symmetries. At high
temperatures the material also presents a cubic symmetry,
belonging to $O_h^1$ space group.

For PZT with $x \leq 0.05$ the structure at room temperature is
orthorhombic $(C_{2v}^8)$. Increasing the temperature it undergoes
two phase transitions: antiferroelectric orthorhombic
$\rightarrow$ high temperature ferroelectric rhombohedral
$\rightarrow$ paraelectric cubic. For $Ti$ concentrations $0.05
\leq x \leq 0.48$ the material, at room temperature, has a
rhombohedral symmetry and can present two phases: (a) low
temperature ferroelectric rhombohedral phase, $F_R(LT)$, $0.05
\leq x \leq 0.37$; (b) high temperature ferroelectric
rhombohedral, $F_R(HT)$, $0.37 \leq x \leq 0.48$. Increasing the
temperature and starting from $T=300 K$ into $F_R(LT)$ phase, it
is observed a phase transition to $F_R(HT)$ and a second phase
transition to a paraelectric cubic phase. By other side, if $x
\geq 0.48$ the PZT room temperature phase has tetragonal symmetry
and increasing the temperature a phase transition to a
paraelectric cubic phase is observed. A complete
temperature-concentration diagram is given in Ref. 13.

In order to construct the concentration-temperature-pressure
diagram for PZT, some investigations have been reported. Bauerle
et. al.\cite{pinzuk} have studied $PbTi_{0.10}Zr_{0.90}O_3$ by
Raman spectroscopy with pressures up to 6.85 $GPa$. They showed
that the material undergoes a phase transition at 0.57 $GPa$ from
the initial room temperature-room pressure ($F_R(LT)$) to a
high-temperature rhombohedral phase. Between 0.8 and 0.91 $GPa$,
$PbTi_{0.10}Zr_{0.90}O_3$ goes to the orthorhombic
antiferroelectric phase and between 3.97 and 4.2 $GPa$ a new phase
is reached, with a symmetry higher than that of the
antiferroelectric phase.

In recent high-pressure works, Furuta et. al. showed by X-ray
\cite{endo} and Raman \cite{furuta} measurements that $PbZrO_3$
polycrystalline fine-powder sample undergoes two phase transitions
up to 30 $GPa$: from the antiferroelectric phase to an
orthorhombic phase I' at 2.3 $GPa$ and from an orthorhombic phase
I' to an orthorhombic phase I'' at 17.5 $GPa$.

The purpose of the present work is to present an investigation of
the $PbZr_{0.94}Ti_{0.06}O_3$ by high-pressure Raman spectroscopy,
in the pressure range between 0.0 and 3.7 $GPa$. The results
presented in this work yield new informations about the PZT
pressure-concentration diagram and indicate that the sequence of
pressure induced phase transitions in $PbZr_{0.94}Ti_{0.06}O_3$ is
very different from those of $PbZr_{0.90}Ti_{0.10}O_3$, although
the initial (atmospheric) phase are the same for both materials.

\section{Experimental}

The samples were prepared by the usual ceramic technique from 99.9
$\%$ pure reagent grade PbO, $ZrO_2$ and $TiO_2$ oxides.
$Pb(Zr_{1-x}Ti_x)O_3$ powder at 0.86 $<$ x $<$ 0.98 mol $\%$ were
prepared. A mixture of the starting powders and distilled water
was milled in a ball mill during 3.5 hours for homogenization. The
mixture was calcined at 850 $^oC$ for 2.5 h. The fine powder
obtained was pressed into disks of  20 mm diameter and 3 mm
thickness at 4,9 x $10^5$ Pa. Finally, the disks were fired in
covered alumina crucibles at a sintering temperature of 1250 $^oC$
for 4 h. The atmosphere was enriched in PbO vapor using $PbZrO_3$
powder around the disks to prevent significant volatilization of
PbO. PZT samples were microstructurally characterized by X-ray
diffraction, using a RIGAKU difractometer, model DMAXB, with a
$CuK\alpha$ radiation (40 KV and 25 mA). Measurements were
performed on the PZT powder using step mode (0.01 degrees/step)
with 5 seconds for counting time per step in the range of 15 to 60
degrees.

Small pieces of samples with thickness of 100 $\mu$m were cut
under a microscope to load the pressure cell. The pressure cell
was a diamond anvil cell (DAC) with 4:1 methanol-ethanol as
pressure-transmitting fluid. The pressure was measured by the
frequency shift of the ruby luminescence lines. Excitation with
514.5 nm radiation from an argon ion laser working at 10 mW was
employed. Backscattering configuration was the chosen geometry of
this work. Raman spectra were obtained at room temperature by a
Jobin Yvon Triplemate 64000 spectrometer equipped with
$N_2$-cooled charge-coupled device (CCD) detection system. The
slits were set for a 2 $cm^{-1}$ spectral resolution and a
microscope lenses with $f=20,5$ $mm$ was used to focus the laser
on the sample.

\section{Results and Discussion}

The PZT samples used in our experiments had Ti concentrations of 6
$\%$. At room temperature the structure has a rhombohedral (low
temperature) symmetry, as observed by X-ray diffraction. In Fig. 1
we show Raman spectra of PZT taken at various pressures in the
frequency range 25 to 200 $cm^{-1}$. This frequency region covers
mainly the lattice modes of the material. The room pressure (0.0
$GPa$) spectrum, typical of the rhombohedral phase, displays three
bands for frequencies lower than 200 $cm^{-1}$ at 52, 61 and 128
$cm^{-1}$. Fig. 2 shows Raman spectra of PZT taken at different
pressures in the frequency range 150 - 750 $cm^{-1}$. In this
spectral region it is expected to be observed some internal modes
of certain groups of the material. In the 0.0 $GPa$ spectrum it is
observed bands at 207, 241 (Zr-O bending), 282 and 328 ($ZrO_3$
torsions), 535 (Zr-O stretching) and 670 $cm^{-1}$. The
assignments of these bands were made based on the works in
$PbZrO_3$ single crystals \cite{pasto}.

Increasing the pressure the Raman spectra remain qualitatively the
same of the room temperature pressure up to 0.3 $GPa$. This means
that the rhombohedral phase is stable up to that pressure.
However, in the spectrum taken at0.3 $GPa$, completely different
bands are observed. Fig. 1 shows that in the0.3$GPa$ spectrum, new
bands appear at 36, 45 and 56 $cm^ {-1}$. At the same time a band
with frequency of 61 $cm^ {-1}$ in the 0.26 $GPa$ spectrum
disappears. Fig. 2 also shows drastic changes in the Raman spectra
between 0.26 and 0.3 $GPa$. The 207 $cm^ {-1}$ band grows in
intensity; the 241 $cm^{-1}$ jumps to lower frequency (235 $cm^
{-1}$) and decreases in intensity; the 328 $cm^ {-1}$ band
decreases in intensity and a new band, at 346 $cm^{-1}$, appears.
The discontinuities both in the number of bands and in the band
frequencies and the changes in the band's intensities point to a
phase transition from the rhombohedral structure to a different
one. In Fig. 3 we present the frequencies of all bands observed in
our PZT Raman spetra as a function of the pressure in the range
0.0 to 3.7 $GPa$, where the discontinuites outlined above can be
clearly seen. We remember that for $PbZr_{0.9}Ti_{0.1}O_3$, this
material undergoes three different phase transitions from a
$F_R(LT)$ to a rhombohedral phase, $F_R(HT)$, from that to an
orthorhombic phase and, finally, to a phase of high symmetry
\cite{pinzuk}.

It should be observed that the initial phase of the material
studied in this work is the same of $PbZr_{0.90}Ti_{0.10}O_3$.
However, are the sequence of phase transition in
$PbZr_{0.90}Ti_{0.10}O_3$ and $PbZr_{0.94}Ti_{0.06}O_3$ the same?
In order to determine the sequence of phase transitions of
$PbZr_{0.94}Ti_{0.06}O_3$ one has to search the various phases
presented by the material. Let us now try to provide the correct
structure of the new phase observed in the PZT for P $\ge$ 0.3
$GPa$. The standard process would be to perform X-ray measurements
simultaneously with our Raman measurements. However, our
experimental setup does not allow us to do X-ray measurements at
high-pressure. Another possibility it is to compare the Raman
spectra taken in the new phase with the Raman spectra of
$PbZr_{1-x}Ti_{x}O_3$ with different Ti concentrations, but with
well-known phases.

\begin{figure}
\centerline{\epsfig{file=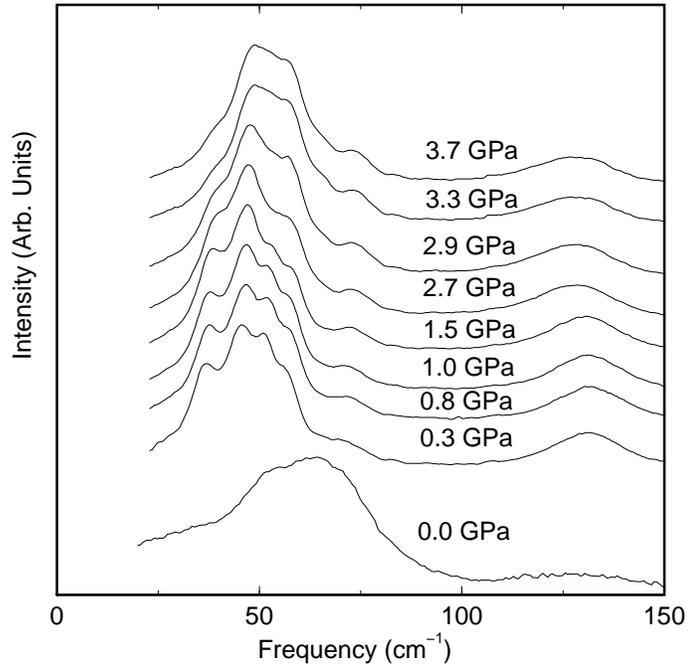,width=30pc}} \caption{Raman
spectra of $PbZr_{0.94}Ti_{0.06}O_3$ at various pressures in the
low frequency region.}
\end{figure}

\begin{figure}
\centerline{\epsfig{file=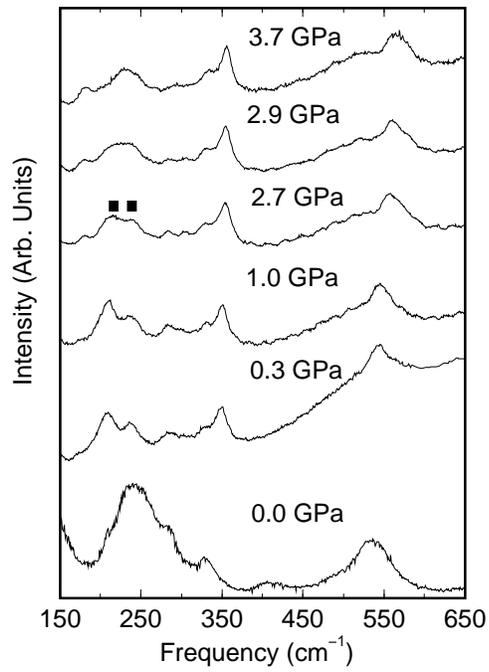,width=30pc}} \caption{Raman
spectra of $PbZr_{0.94}Ti_{0.06}O_3$ at various pressures in the
high frequency region.}
\end{figure}

We performed a series of both Raman and X-ray measurements in
$PbZr_{1-x}Ti_{x}O_3$ with different Ti concentrations at room
temperature and atmospheric pressure. The Raman spectra of Fig. 1
and Fig. 2 obtained at pressure P=0.3 $GPa$ (the new phase) is the
same that we observed for Ti concentrations of 0.02 and 0.04; in
those cases X-ray diffraction showed that at room temperature, the
phase of the material is orthorhombic, as pointed by other works
in the literature. Also, the Raman spectrum is the same reported
to $PbZrO_3$ single crystal that belongs to orthorhombic structure
\cite{pasto}. We can conclude that the new phase presented by
$PbZr_{0.94}Ti_{0.06}O_3$ is an orthorhombic one and the first
phase transitions we are observing is from a rhombohedral
$F_R(LT)$ to an orthorhombic phase.

Fig. 3 shows that this new phase is stable over a large range of
pressure; in the figure the two dashed lines limit the
orthorhombic phase region from 0.3 to 2.7 $GPa$. Some features of
the orthorhombic phase Raman spectra, as a function of pressure,
can be described as follows: (a) the frequencies of bands change
only slightly; (b) the band at 36 and 50 $cm^{-1}$ decrease slowly
in intensity up to 2.7 $GPa$; (c) the band at 70 $cm^{-1}$
increases slowly in intensity as pressure is increased.

\begin{figure}
\centerline{\epsfig{file=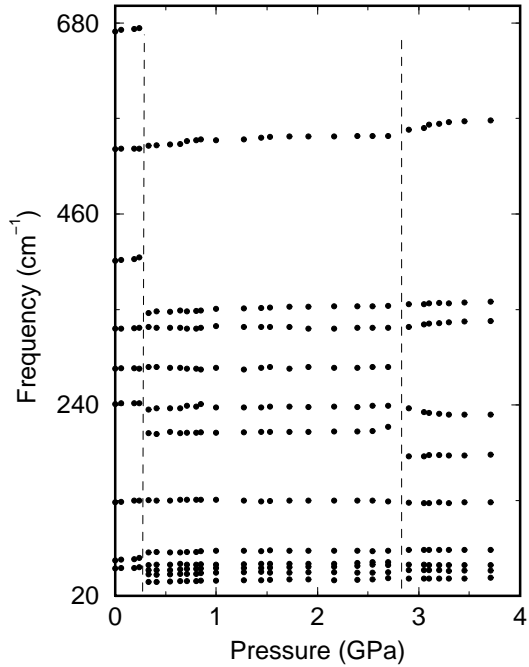,width=30pc}}
\caption{Pressure dependence of the Raman bands of
$PbZr_{0.94}Ti_{0.06}O_3$ at room temperature. The dahsed lines
represent the phase transition pressures}
\end{figure}

However, for pressures higher than 2.9 $GPa$, different Raman
spectra are observed. Particularly clear is the disappearing of
both a band in the lattice region of spectra (Fig. 1) and two
bands marked by squares in Fig. 2. Fig. 3 also shows these
discontinuities in the frequency of Raman bands. For frequencies
lower than 150 $cm^{-1}$ six bands are observed in the spectra for
pressure ranging from 0.3 to 2.7 GPa. For pressures higher than
2.7 GPa just five bands are observed. The band whose intensity
goes to zero for P $\ge$ 2.7 GPa was found in the 0.3 GPa spectrum
at 56 $cm^{-1}$. As pressure increases, its frequency increases
and its intensity decreases. This intensity decreasing is slow,
associated with the continuous change from the intermediate
structure to the structure of higher pressure (P $\ge$ 2.7 GPa).

The Raman spectra for frequencies higher than 100 $cm^{-1}$ at 3.3
$GPa$ can all be identified on the basis of previous work
performed on $PbZrO_3$ \cite{furuta}. If we compare the Raman
spectrum at 3.3 $GPa$ in Fig. 3, we observe that it is similar to
the spectrum of $PbZrO_3$ at 2.3 $GPa$. At this pressure $PbZrO_3$
belongs to an orthorhombic phase(I') \cite{furuta} , different
from the original orthorhombic phase(I), where dielectric
hysteresis measurements show to be an antiferroelectric phase
\cite{endo} . This indicates that in $PbZr_{0.94}Ti_{0.06}O_3$,
there is a phase transition from an orthorhombic phase(I) to an
orthorhombic phase(I') at about 2.9 $GPa$.

In summary, the results Raman measurements presented here have
showed that $PbZr_{0.94}Ti_{0.06}O_3$ undergoes two different
phase transitions up to 3.7 $GPa$: $rhombohedral(LT)\stackrel{0.3
GPa}{\rightarrow} orthorhombic(I) \stackrel{2.9
GPa}{\rightarrow}orthorhombic(I')$. These results point to the
richness of concentration-pressure phase diagram because the
sequence of phase transitions observed are very different from the
$PbZr_{0.90}Ti_{0.10}O_3$ \cite{pinzuk} and $PbZrO_3$
\cite{furuta}.

\bigskip

{\bf Acknowledgments}: A. G. Souza Filho is grateful for Funda\c
{c}\~ao Cearense de Amparo a Pesquisa (FUNCAP) for fellowship
program. The authors would like to thank CNPq (Conselho Nacional
de Desenvolvimento Cient\'{\i}fico e Tecnol\'ogico) and FUNCAP (
Funda\c {c}\~ao Cearense de Amparo a Pesquisa) Brazilian funding
agencies for financial support.


\begin{references}


\bibitem{hz}
H. Zhang, S. Leppavuori, and P. Karjalainen, J. Appl. Phys., 77,
(1995) 2691.

\bibitem{pig}
A. Pignolet, L. Wang, M. Proctor, F. Levy, and P.E. Schmid, J.
Appl. Phys., 74, (1993) 6625.

\bibitem{kata}
H. Fujishita and S. Katano, J. Phys. Soc. Jpn, 66, (1997) 3484.

\bibitem{z}
Z. Ujma, J. Handerek, H. Hassan, G.E. Kugel, and M. Pawelezyk, J.
Phys: Condens. Matter, 7, (1995) 895.


\bibitem{ujma}
J. Handerek and Z. Ujma, J. Phys: Condens. Matter., 7, (1995)
1721.

\bibitem{becker}
I. El-Harrad, P. Becker, C. Carbatos-Nedelec, J. Handerek, Z. Ujma
and D. Dimytrov, J. Appl. Phys., 78, (1995) 5581.

\bibitem{db}
D. Baurele, Y. Yacoby, and W. Ricther, Solid State Commun., 14,
(1974) 1137.

\bibitem{dca}
D.C. Agrawal, S.B. Majumder, Y.N. Mohapatra, S. Sathaiah, H.D.
Bist, R.S. Katiyar, E. Ching-Prado,and A. Reynes, J. Raman
Spectroscopy, 24, (1993) 459.

\bibitem{kr}
K. Rolender, G. E. Kugel, M.D. Fontana, J. Handerek, S. Lahlon and
C. Carbatos-Nedelec, J. Phys: Condens. Matter, 1 (1989) 2257.

\bibitem{burns}
G. Burns, and B.A. Scott, Phys. Rev Lett., 25, (1970) 167.

\bibitem{scott}
G. Burns, and B.A. Scott, Phys. Rev Lett., 25, (1970) 1191.

\bibitem{meng}
J.F. Meng, R.S. Katiyar, G.T. Zou, and X.H. Wang, phys. stat.
sol.(a), 164, (1997) 851.

\bibitem{bj}
B. Jaffe, W.R. Cook, and H. Jaffe, Piezoelectric Ceramic, Academic
Press, New York, 1971, pg. 139.

\bibitem{pinzuk}
D. Bauerle, W.B. Holzapel, A. Pinczuk and Y. Yacoby, phys. stat.
sol., 83, (1977) 99.

\bibitem{endo}
Y. Kobayashi, S. Endo, L.C. Ming, K. Deguchi, T. Ashida and H.
Fujishita, J. Phys. Chem. Solids, 60, (1999) 57.

\bibitem{furuta}
H. Furuta, S. Endo, L.C. Ming and H. Fujishita, J. Phys. Chem.
Solids, 60, (1999) 65.

\bibitem{pasto}
A.E. Pasto, and R.E. Condrate, J. Am. Ceram. Soc., 56, (1973) 436.

\end{references}
\end{document}